\documentclass[twocolumn,showpacs,preprintnumbers,amsmath,amssymb,aps,pra,groupedaddress]{revtex4-2}
\usepackage{multirow,diagbox}  
\usepackage{siunitx}
\usepackage[colorlinks,citecolor=blue,linkcolor=red,hyperindex,CJKbookmarks]{hyperref}
\usepackage{amsmath}
\usepackage{graphicx}
\usepackage{dcolumn}
\usepackage{subfigure}
\usepackage{mathrsfs}
\usepackage{amssymb}
\usepackage{amsfonts}
\usepackage{makecell}
\usepackage{color}
\begin{document}
\draft
\title{Critical quantum sensing based on the Jaynes-Cummings model \\
	with a squeezing drive}
\author{Jia-Hao L\"{u}}
\author{Wen Ning, Xin Zhu, Fan Wu}
\author{Li-Tuo Shen}\thanks{E-mail: lituoshen@yeah.net}
\author{Zhen-Biao Yang}\thanks{E-mail: zbyang@fzu.edu.cn}
\author{Shi-Biao Zheng}
\address{Fujian Key Laboratory of Quantum Information and Quantum\\
Optics, College\\
of Physics and Information Engineering, Fuzhou University, Fuzhou, Fujian
350108, China}
\date{\today }

\begin{abstract}
	Quantum sensing improves the accuracy of measurements of relevant parameters by exploiting the unique properties of quantum systems. The divergent susceptibility of physical systems near a critical point for quantum phase transition enables criticality-enhanced quantum sensing. The quantum Rabi model (QRM), composed of a single qubit coupled to a single bosonic field, represents a good candidate for realizing such critical enhancement for its simplicity, but it is experimentally challenging to achieve the ultrastrong qubit-field coupling required to realize the critical phenomena. In this work, we explore an alternative to construct the analog of the QRM for the sensing, exploiting the criticality appearing in the Jaynes-Cummings (JC) model whose bosonic field is parametrically driven, not necessitating the ultrastrong coupling condition thus to some extent relaxing the requirement for the practical implementation.
\end{abstract}
\pacs{~}

\vskip 0.5cm

\narrowtext
\maketitle
\section{INTRODUCTION}

Quantum sensing makes use of the unique properties, such as quantum coherence or entanglement, in quantum physics, to improve the sensitivity of measurement \cite{chuangan1,chuangan2,chuangan3,chuangan4,chuangan5}. Recently, it was realized that high-precision quantum metrology could be realized by encoding the signal in a quantum system near its critical point, where it exhibits an ultrasensitive response to a tiny change in the relevant parameter \cite{ref15,ref16,ref17,ref18,ref19,ref20,ref20.5,ref21,ref22,ref23,ref24}. So far, two complementary approaches have been proposed for criticality-enhanced quantum sensing based on equilibrium or steady state response to the signal change. One approach is to exploit the equilibrium properties of critical systems, whose Hamiltonians are slowly varied to ensure the system to remain in the ground states \cite{pingheng1,ref20.5,pingheng2,pingheng3}. For an open system under the interplay between the Hamiltonian dynamics and the dissipative process, its nonequilibrium behavior in a steady state near a dissipative-driven phase transition can also be utilized for critical sensing \cite{pingheng3,wentai1,wentai2,wentai3,wentai4}. Both methods suffer from the relatively long-time evolution required for satisfying the adiabaticity \cite{juereyaoqiu}, restricting their practical implementations. To overcome this problem, a dynamical method was recently proposed \cite{ref21}, where the signal is encoded in a dynamically-evolved feature, without requirement of adiabatic or slow quench condition. Furthermore, this method is quite general, working even for a mixed state.

The quantum Rabi model (QRM) \cite{ref1,ref2,ref3,ref4} is an ideal system for realizing these approaches. The QRM, which describes the interaction between a two-level atom and a single-mode bosonic field, has been widely investigated in quantum information and technology. Under the condition that the ratio between the atomic transition frequency and the field frequency tends to  infinity, such a model can undergo a quantum phase transition of the normal phase to the superradiant phase, featuring a sudden increase of the photon number at a critical point of the coupling-to-frequency ratio \cite{ref5,ref66,ref7,ref8,ref9,ref10,ref11,ref12,ref13,ref14}. However, the criticality accompanied by the quantum phase transition realized in the QRM requires the ultra-strong coupling between the two-level system and the single-mode bosonic field \cite{coupling1,coupling2}, so that the counter-rotating-wave terms play a significant role, which represents an experimental challenge. Another restriction originates from the so-called no-go theorem \cite{nogo1,nogo2,nogo3}, which states the neglected \textit{$A^2$} term would prohibit the superradiant phase transition. The dynamics of the QRM has been simulated on a variety of platforms, such as superconducting circuit systems \cite{coupling1,exp2} and the trapped ions \cite{coupling2}. Very recently, the superradiant phase transition was observed in an effective QRM, composed of a superconducting qubit and with a microwave photonic field stored in a resonator \cite{zhengrihua}, whose interaction was engineered with two deliberately tailored longitudinal modulations and a transverse drive. However, experimental demonstration of criticality-enhanced quantum sensing in such a system remains elusive.

We here explore the criticality-enhanced sensing with a squeezed JC model (SJCM), where the bosonic field is subjected to a parametric squeezing drive. This parametric driving transforms the rotating-wave coupling of the JC model into a combination of the rotating and counter-rotating couplings, realizing an isomorphism of the QRM, where the system frequencies are replaced by their detunings from the driving \cite{ref261}. Such an isomorphism offers the possibility to explore light-matter interactions in the ultra-strong coupling regime based on the JCM, to bypass the no-go theorem to access the superradiant phase transition predicted in the QRM \cite{ref26}, and to harness the associated critical phenomena for quantum technology.

With this quantum critical dynamics, a divergent behavior emerges in the quantum Fisher information (QFI) under certain conditions when it approaches the critical point, allowing for a ultrasensitive measurement of the relevant observable. The results of our study show that such a scheme can achieve the precision close to the quantum Cram\'{e}r-Rao bound \cite{cr1,cr2,cr3}, which is given by the Cram\'{e}r-Rao inequality $\langle\Delta^2\hat\theta\rangle\ge\frac{1}{\nu\mathcal{F}}$, where $\langle\Delta^2\hat\theta\rangle$ is the variance of observable $\hat\theta$, $\nu$ is the amount of data and $\mathcal{F}$ represents the QFI. The QFI gives an absolute lower bound on the measurement of an input state, independent of the measurement method, and is equivalent to the inverse variance of the measurement, which provides great convenience to reflect the achievable measurement precision. We also give the analytical result of the QFI for the correlated quantum dynamics, and propose a scheme for experimental implementation of the model. 

\section{The QFI in the critical quantum systems}
The QFI about the parameter $\alpha$ can be expressed as $\mathcal{F}_\alpha=4{\rm Var}[h_\alpha]_{\vert\Phi\rangle}$, 
where $h_\alpha=-i(\partial_\alpha U^\dag_\alpha) U_\alpha=iU^\dag_\alpha (\partial_\alpha U_\alpha)$ and ${\rm Var}[h_\alpha]_{\vert\Phi\rangle}$ is the variance of $h_\alpha$ with respect to the initial state $\vert\Phi\rangle$ \cite{cr5}.
 We consider the Hamiltonian $\hat H_\alpha=\hat H_0+\alpha \hat H_1$, which satisfies the relation \cite{cr4}
 \begin{eqnarray}
 	[\hat H_\alpha,\hat \Upsilon]=\sqrt{\Delta}\hat \Upsilon, \label{114}
 \end{eqnarray}%
where $\hat{\Upsilon}=i\sqrt{\Delta}\hat M-\hat N$ with $\hat M=-i[\hat H_0,\hat H_1]$, $\hat N=-[\hat H_\alpha,[\hat H_0,\hat H_1]]$. $\Delta$ depends on the parameter $\alpha$.  This kind of Hamiltonian contributes to the equally spaced gap $\epsilon\sim\sqrt\Delta$ for $\Delta>0$ and becomes imaginary if $\Delta<0$ \cite{ref21}, which shows that the quantum phase transition behaviors occur at the critical point $\alpha=\alpha_c$ defined by $\Delta=0$. Besides,  $h_\alpha$ can be expanded as 
 \begin{eqnarray}
	h_\alpha=-i\sum_{n=0}^{\infty}\frac{(it)^{n+1}}{(n+1)!}[\hat H_\alpha,\hat H_1]_n,
\end{eqnarray}%
where $[\hat H_\alpha,\hat H_1]_n=[\hat H_\alpha,[\hat H_\alpha,\hat H_1]_{n-1}]$ and $[\hat H_\alpha,\hat H_1]_0=\hat H_1$. As shown above, we can express the commutation relations in terms of $\hat M$ and $\hat N$ and could get the following expression:
{\small\begin{eqnarray}
h_\alpha=\hat H_1t+\frac{\cos{(\sqrt\Delta t)}-1}{\Delta}\hat M-\frac{\sin{(\sqrt\Delta t)}-\sqrt\Delta t}{\Delta^{\frac{3}{2}}}\hat N.
\end{eqnarray}}
It shows that  as $\Delta\to 0$, $h_\alpha$ exhibits a divergent behavior. Close to this point, the term proportional to $\Delta^{-3/2} $ is dominant. And, the QFI can be expressed as 

 \begin{eqnarray}
	\mathcal F_\alpha(t)\simeq4\frac{[\sin{(\sqrt\Delta t)}-\sqrt\Delta t]^2}{\Delta^3}{\rm Var}[\hat N]_{\vert\Phi\rangle} \label{113}.
\end{eqnarray}%
If $\sqrt\Delta t\simeq\mathcal O(1)$, $\mathcal F_\alpha(t)$ is divergent at $\Delta=0$ and scales with  $\Delta^{-3}$. As was pointed out in Ref. \cite{ref21}, such a scaling of the QFI holds for general initial states  $\vert\Phi\rangle$ provided $Var[\hat N]_{\vert\Phi\rangle}\simeq\mathcal O(1)$ or even more general mixed states.

\section{quantum sensing with the SJCM}
  We here consider the system composed of a JC model with a squeezing drive to the bosonic field. The Hamiltonian can be written as $(\hbar=1)$ 
\begin{eqnarray}
	\hat H=\omega \hat a^{\dagger}\hat a+\frac{\Omega}{2}\hat \sigma_z+\lambda(\hat a^{\dagger}\hat \sigma_-+\hat a\hat\sigma_+)-G(\hat a^2+\hat a^{\dagger2}), \label{112}
\end{eqnarray}%
where $\Omega$ is the frequency of the two-level system, $\hat\sigma_{x,y,z}$ are the Pauli operators and  $\hat\sigma_\pm=(\hat\sigma_x \pm i\hat\sigma_y)/2$, $\hat a^{\dagger}(\hat a)$ is the creation (annihilation) operator of the bosonic field with the frequency $\omega$. G is the driving strength \cite{twophoton1,twophoton2,twophoton3} and $\lambda$ is the coupling strength between the two-level system and the bosonic field.

 Under the condition $\Omega/\omega\to \infty$ and $\Omega/G\to \infty$, we can use the Schrieffer-Wolff transformation $\hat H_S=e^{-\hat S}\hat He^{\hat S}$ with an anti-Hermitian and block-off diagonal operator $\hat S={\lambda}(\hat a^{\dagger}\hat\sigma_--\hat a\hat\sigma_+)/{\Omega}$ to remove the interaction term related to $\lambda$ \cite{ref5}. The low-energy Hamiltonian for the effective normal-phase can be obtained as $\hat H^{\downarrow}_{np}=(\omega-\lambda^2/{\Omega})\hat a^{\dagger}\hat a-G(\hat a^{\dagger2}+\hat a^2)-{\Omega}/{2} $. We diagonalize this effective Hamiltonian and introduce a squeezed transformation $\hat H^{'}_{np}=e^{{r}(\hat a^{\dagger2}-\hat a^2)/2}\hat H^{\downarrow}_{np}e^{-{r}(\hat a^{\dagger2}-\hat a^2)/2}$. This implies that if $r=\frac{1}{4}\ln{\frac{\omega-{\lambda^2}/{\Omega}-2G}{\omega-{\lambda^2}/{\Omega}+2G}} $, the phase transition occurs at the critical point $g=1$ with $g=\lambda/\sqrt{\Omega(\omega-2G)}$ \cite{sjcm}. Additionally, the Hamiltonian $\hat  H^{'}_{np}$ satisfies the relation in Eq. (\ref{114}) with $\Delta=16\alpha(\alpha+2G)$ and $\alpha=(\omega-2G)(1-g^2)/2$, which indicates that the nonanalytic behaviors would take place at the critical point $g=1$ for $\Delta=0$. Now we show the ultimate precision of the quantum parameter estimation defined by the Cram\'{e}r-Rao bound. The measurement precision of the parameter $g$ can be given by QFI in Eq. (\ref{113}) as 
  
  \begin{eqnarray}
  \mathcal{F}_g(t)&\simeq&1024(\omega-2G)^2G^2g^{2}(\alpha+2G)^2  \nonumber\\
  &&\times\frac{[\sin(\sqrt{\Delta}t)-\sqrt{\Delta}t]^2}{\Delta^3}{\rm Var}[\hat {P}^2]_{\left\vert \varphi\right\rangle}, \label{111}
  \end{eqnarray}%
 where $\hat {P}=i(\hat a^{\dag}-\hat a)/\sqrt2$ is the momentum operator and $\left\vert \varphi\right\rangle$ is the initial state of the bosonic field. It is obvious that $\mathcal{F}_g(t)\to \infty$ when $g\to1$, indicating we can rely on the critical dynamics to estimate the precision of the relevant parameter associated with $g$. The quantum sensing can be realized by encoding the physical quantity of interest either in the qubit or in the field. We will analyze the performances of the two different encoding schemes in detail.
 \section{Encoding schemes}
 We first investigate the performance of the quantum sensor that works by encoding the signal in one of the quadratures of the field. For convenience, here we set the bosonic field to be initially in the state $\left\vert \varphi\right\rangle=(\left\vert 0\right\rangle+i\left\vert 1\right\rangle)/\sqrt2$ and the two-level system in its ground state. After an interaction time $t$, the expectation value and variance of the field quadrature $\hat {X}=(\hat a+\hat a^{\dag})/\sqrt2$ evolve as
 
\begin{eqnarray}
\langle \hat{X}\rangle_t=2\sqrt2(\alpha+2G)\Delta^{-\frac{1}{2}}\sin(\sqrt{\Delta}t/2),           \label{124}
\end{eqnarray}
and
\begin{eqnarray}
(\Delta\hat{X})^2&=&\cos^2(\sqrt{\Delta}t/2)  \nonumber\\
&+&8(\alpha+2G)^2\Delta^{-1}\sin^2(\sqrt{\Delta}t/2) ,   \label{123}
\end{eqnarray}%
respectively (see Appendix \ref{A} for detailed derivation). 
\begin{figure}
	\includegraphics[width=0.5\textwidth]{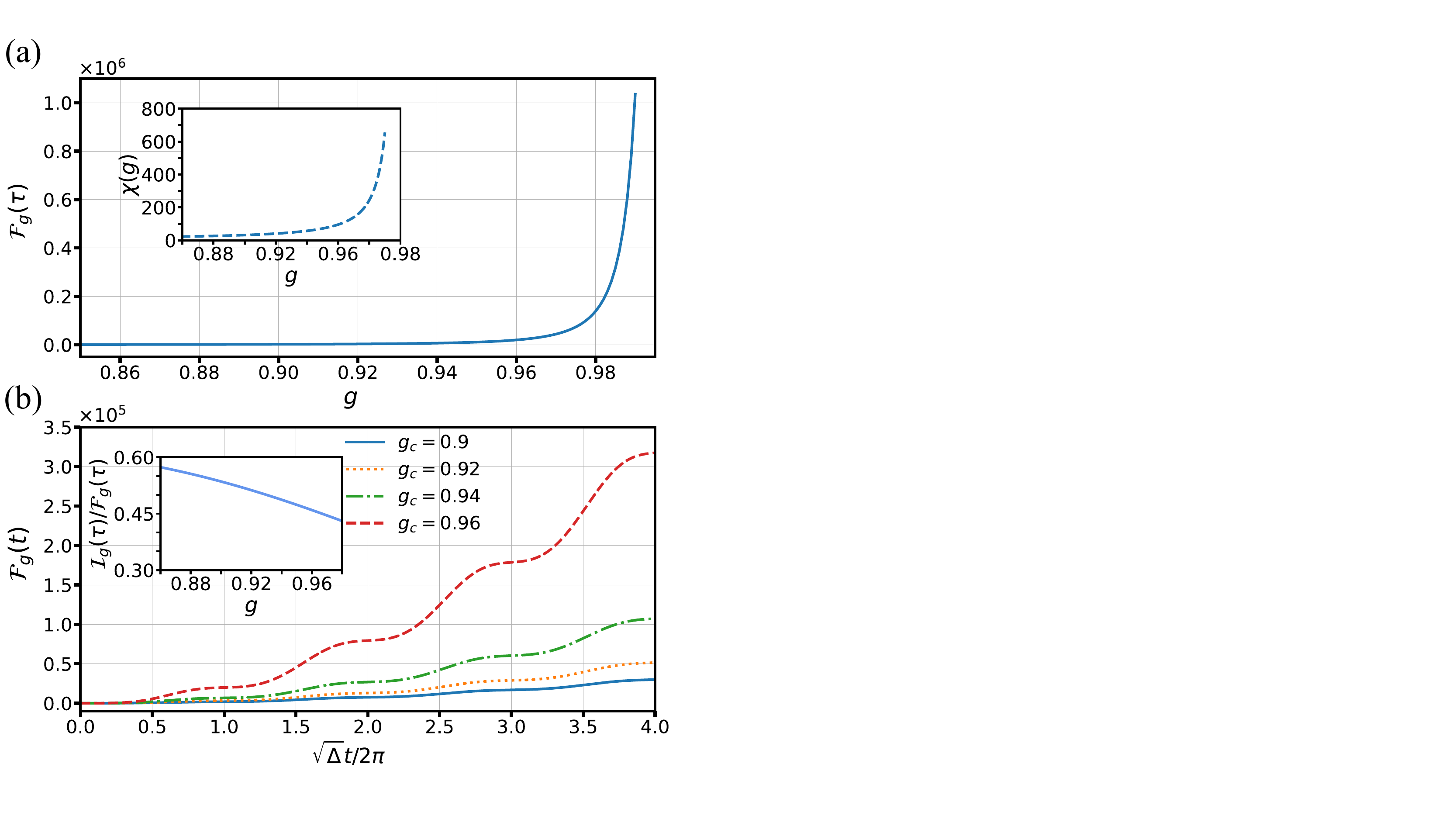}
	\caption{Quantum sensing by homodyne detection of the bosonic field. (a) The QFI $\mathcal{F}_{g}(t)$ as a function of $g$ for an evolution time $\tau=2\pi/\sqrt{\Delta}$. When $g$ is close to the critical point, $\mathcal{F}_{g}(t)$ exhibits a divergent behavior. Inset: After an evolution time $\tau=2\pi/\sqrt{\Delta}$, the susceptibility $\chi_{g}(t)$ as a function of $g$, which exhibits a divergent behavior.  (b) The QFI $\mathcal{F}_{g}(t)$ as a function of the evolution time $t$. Inset: For $\tau=2\pi/\sqrt{\Delta}$, the local maximum of the inverted variance $\mathcal{I}_g(t)$ is of the same order as $\mathcal{F}_g(t)$.}
	\label{aaa}
\end{figure}
 The inverted variance is defined by $\mathcal{I}_g(t)=\chi ^2_g(t)/(\Delta\hat{X})^2$, where $\chi_g(t)=\partial_g\langle\hat{X}\rangle_t $ is the susceptibility of the observable $\langle\hat{X}\rangle_t$ associated with  $g$, and exhibits a divergent feature when $g$ is close to the critical point, as shown in Fig. \ref{aaa}(a). To quantify the precision of the relevant parameter estimation, we should compare  $\mathcal{I}_g(t)$ to $\mathcal{F}_g(t)$, where $\mathcal{F}_g(t)$ represents the absolute lower bound of the measurement defined by quantum Cram\'{e}r-Rao bound \cite{cr2}.
Apparently, the inverted variance reaches its maximum at $\tau_n=2 n\pi/\sqrt{\Delta}$ $(n\in{N}^+)$:

\begin{eqnarray}
	\mathcal{I}_g(\tau_n)&=& 2048n^2 \pi^2 (\omega-2G)^2 \nonumber\\
	& & \times g^{2}(\alpha+G)^2(\alpha+2G)^2\Delta^{-3}.
\end{eqnarray}%
It can be derived from Eq. (\ref{111}) to get the QFI $\mathcal{F}_g(\tau_n)\simeq8n^2\pi^2G^2g^{2} (\omega-2G)^{-1}(\alpha+2G)^{-1}(1-g^2)^{-3}{\rm Var}[\hat {P}^2]_{\left\vert \varphi\right\rangle}$. Figure \ref{aaa}(b) shows that  $\mathcal{I}_g(t)$ is close to the QFI, which confirms the feasibility of this protocol. Note that this result does not rely on any particular initial states of the bosonic field. 

Criticality-enhanced quantum sensing can also be realized by encoding the signal in the observables of the qubit. To illustrate the idea, we suppose that the qubit is initially prepared in $\left\vert q\right \rangle=c_1\left\vert \uparrow\right\rangle+c_2\left\vert \downarrow\right\rangle$ and the bosonic field in $\left\vert \varphi\right\rangle=\left\vert 0\right\rangle$. With this initial state, the Bloch vector $\hat\sigma_x$ evolves as  $\langle\hat\sigma_x\rangle=2Re[c^*_1c_2\langle\varphi\vert u^{\dag}_{\uparrow}u_{\downarrow}\vert\varphi\rangle]$, where $u_{\sigma}=e^{-i\hat H_\sigma t}$ is the evolution operator of the bosonic field when the two-level system is in the state $\vert\sigma\rangle$ with $\sigma = $ $\uparrow$ ( or $\downarrow$). Now if $2c^*_1c_2=1$ is chosen, the inverted variance can be simplified as 
  \begin{eqnarray}
  	\mathcal{I}_g=\frac{(\partial_g\langle\hat\sigma_x\rangle)^2}{(\Delta\hat\sigma_x)^2}=\frac{Re[\partial_g \langle\varphi\vert u^{\dag}_{\uparrow}u_{\downarrow}\vert\varphi\rangle]^2}{1-Re[\langle\varphi\vert u^{\dag}_{\uparrow}u_{\downarrow}\vert\varphi\rangle]^2}.
  \end{eqnarray}%
\begin{figure}
	\includegraphics[width=0.5\textwidth]{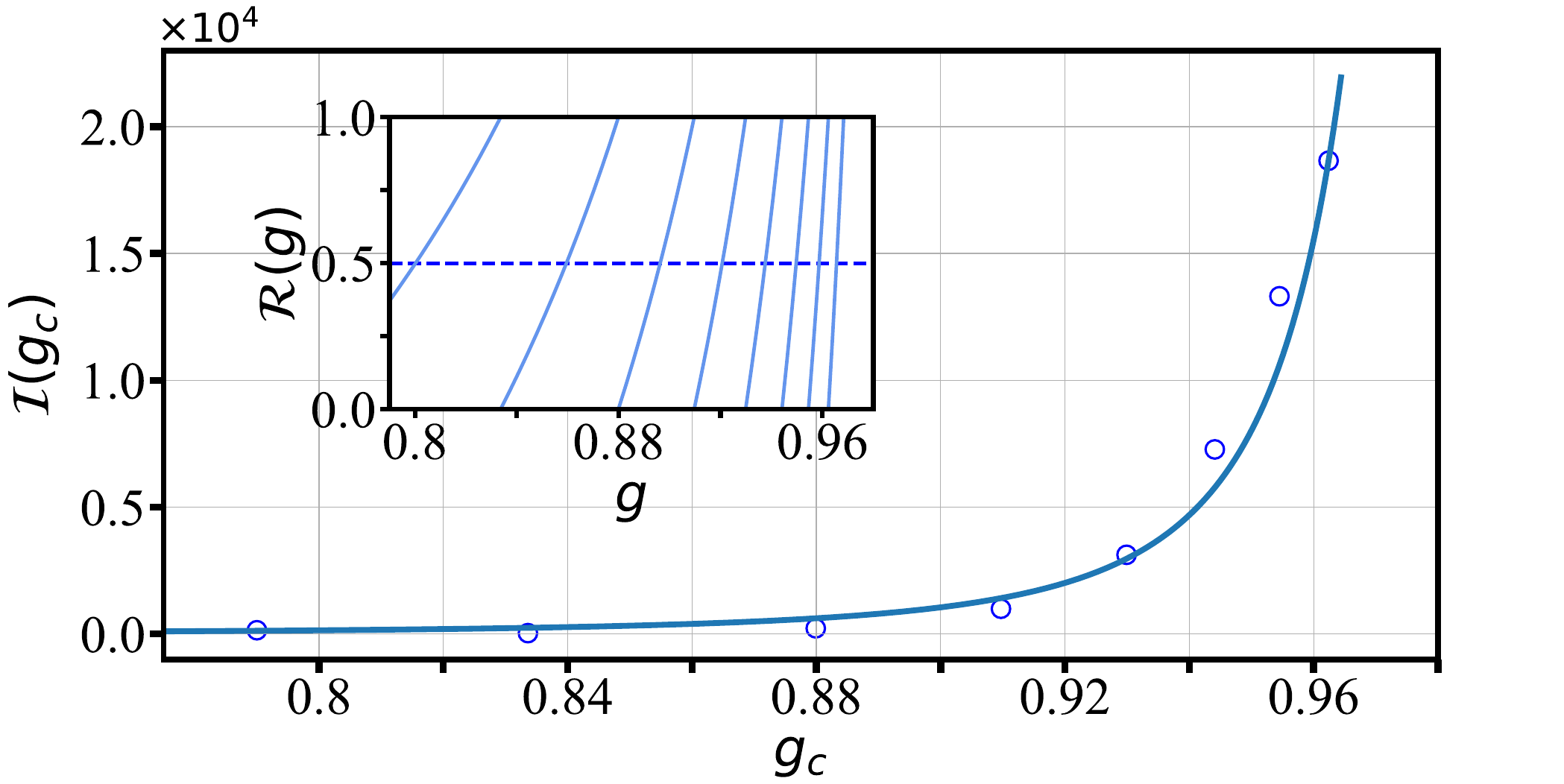}
	\caption{Quantum sensing by local measurement of the two-level system. The circles  denote the working points $g_c$ which are chosen as $\mathcal{R}(g_c)=0.5$. The inverted variance $\mathcal{I}(g_c)$ based on the observable $\langle\hat\sigma_x\rangle$ fits well with $\sim\Delta^{-3}_{g_c}$. The inset shows  $\mathcal{R}(g)$ as a function of $g$. The initial state is $\vert\Phi\rangle=(c_1\left\vert \uparrow\right\rangle+c_2\left\vert \downarrow\right\rangle)\otimes\vert\varphi\rangle$, where $2c^*_\uparrow c_\downarrow=1$ and $\vert\varphi\rangle=\vert 0\rangle$. }
	\label{bbb}
\end{figure}
In Fig. \ref{bbb}, the inverted variance $\mathcal{I}_{g_c}(\tau)=\mathcal{I}_{g}(\tau)\vert_{g=g_c}$ at the working point $g_c$ for $\tau=4\pi/\sqrt\Delta$ is plotted. The result estimates the precision of the parameter $g$ based on the observable $\langle\hat\sigma_x\rangle$. But it should be pointed out that the working point $g_c$ is chosen such that $\mathcal{R}{(g_c)}=0.5$, where $\mathcal{R}_(g)=R(g)-\lfloor R(g)\rfloor$ with $R(g)=\sqrt{\frac{(\omega+{\lambda^2}/{\Omega})^2-4G^2}{(\omega-{\lambda^2}/{\Omega})^2-4G^2}}$ (see Appendix \ref{B}). Figure  \ref{bbb} shows that the inverted variance $\mathcal{I}_{g_c}(\tau)$ exhibits a scaling as $\Delta^{-3}_{g_c}$. This result can be extended to other general initial states, such as the superposition of Fock states.

\section{higher-order corrections}
The analysis shown above is valid in the limitation of $\beta_1=\Omega/\omega\to\infty$ and $\beta_2=\Omega/G\to\infty$, which guarantees the normal-to-superradiant quantum phase transition of the SJCM. However, we can only get the finite values of the ratios when referring to the practical implementations. It is of essential importance to analyze the influence of the higher-order corrections with respect to the Schrieffer-Wolff expansion of the SJCM. 
\begin{figure}
	\includegraphics[width=0.5\textwidth]{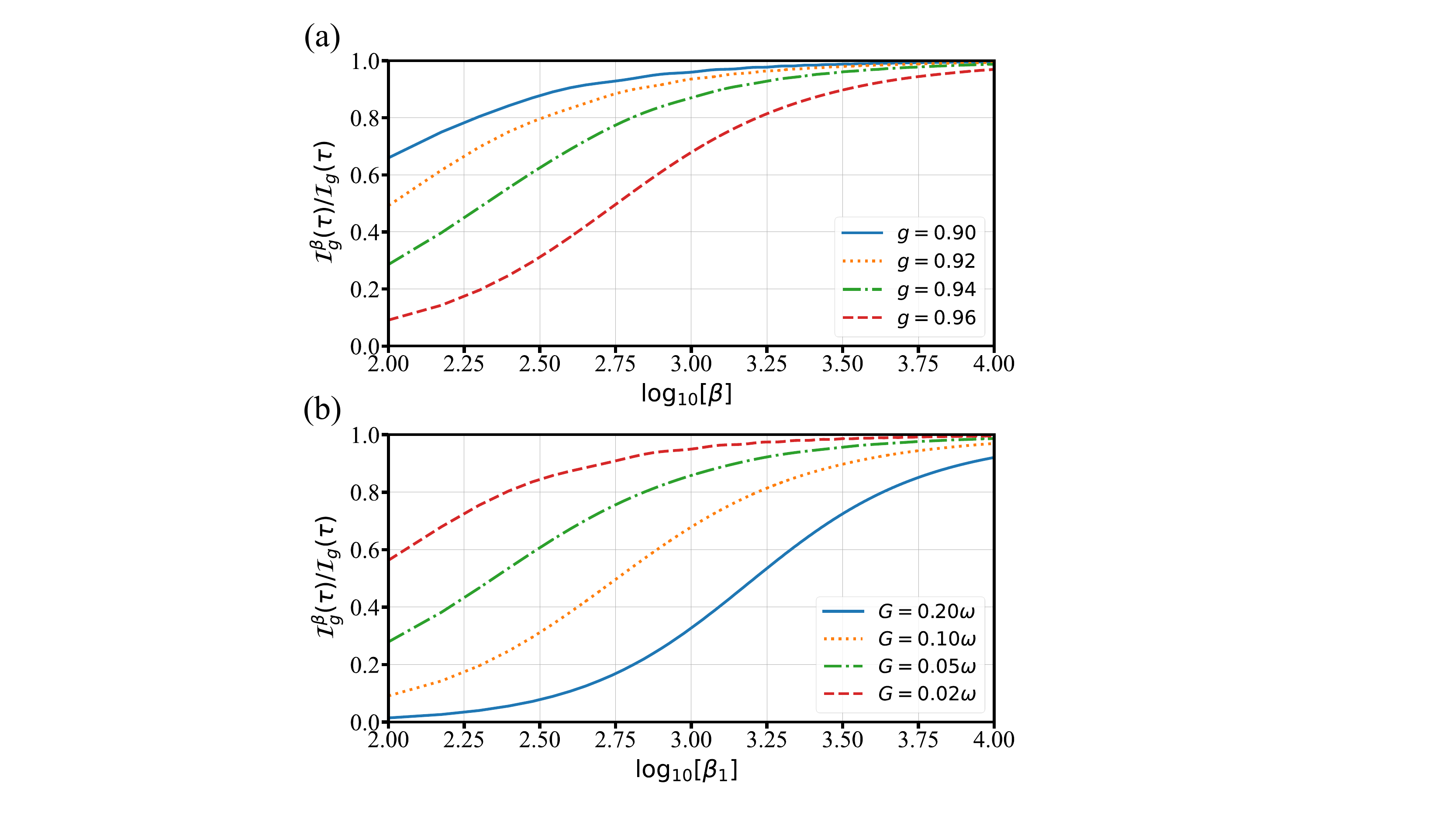}
	\caption{Influence of the finite ratios the qubit frequency to the field frequency and to the squeezing driving strength. (a) Ratio $\mathcal{I}^{\beta}_g(\tau)/\mathcal{I}_g(\tau)$ vs $\beta$ for different values of $g$. In each of the simulated curves  $\beta=\beta_1=10\beta_2$ is set. (b) Ratio $\mathcal{I}^{\beta}_g(\tau)/\mathcal{I}_g(\tau)$ vs $\beta_1$ for different squeezing driving strengths, with $g=0.96$. }
	\label{ccc}
\end{figure}

We correct the expressions Eq. (\ref{124}) and Eq. (\ref{123}) to $\langle\hat{X}\rangle^{exp}_t=\langle\hat{X}\rangle_t+\langle\hat{X}\rangle_{r}$ and $(\Delta \hat{X}^{exp})^2=(\Delta \hat X)^2+(\Delta \hat{X}_{r})^2$. We can rewrite the dynamics of the quadrature in the experimental frame as 
\begin{eqnarray}
\langle\hat{X}\rangle^{exp}_t&=&\langle\Phi\vert e^{i\hat Ht}\hat{X}e^{-i\hat Ht}\vert\Phi\rangle \nonumber \\
&=&\langle\Phi\vert e^{-\hat S_{r}}e^{i\hat H^{\downarrow}_{np}t}e^{-\hat S_{r}}\hat{X}e^{\hat S_{r}}e^{-i\hat H^{\downarrow}_{np}t}e^{\hat S_{r}}\vert\Phi\rangle \nonumber \\
&=&\langle\Phi\vert[1-\mathcal{O}(\beta_i^{{-\frac{1}{2}}})]e^{i\hat H^{\downarrow}_{np}t}\hat{X}[1+\mathcal{O}(\beta_i^{-1})] \nonumber\\
&&\times e^{-i\hat H^{\downarrow}_{np}t}[1+\mathcal{O}(\beta_i^{{-\frac{1}{2}}})]\vert\Phi\rangle,
\end{eqnarray}%
where $\vert\Phi\rangle=\vert\downarrow\rangle_q\otimes\vert\varphi\rangle$ and $\hat S_{r}=\hat S+\frac{\lambda\omega}{{\Omega}^2}(\hat a^\dagger \hat\sigma_--\hat a\hat\sigma_+)+\frac{2G\lambda}{{\Omega}^2}(\hat a^\dagger \hat\sigma_+-\hat a\hat\sigma_-)+\frac{4\lambda^3}{3{\Omega}^3}(\hat a\hat a^\dagger \hat a\hat \sigma_+-\hat a^\dagger \hat a\hat a^\dagger\hat\sigma_-)+\mathcal{O}(\beta_i^{-5/2})$. Obviously, the leading term $\langle\hat{X}\rangle_t=\langle\Phi\vert e^{i\hat H^{\downarrow}_{np}t}\hat{X}e^{-i\hat H^{\downarrow}_{np}t}\vert\Phi\rangle$  is equal to the Eq. (\ref{124}), and  the dominant influence for the correction is on the order of $\beta_i^{{-1/2}}\langle\hat{X}\rangle_t\sim\beta_i^{{-1/2}}\Delta^{-1/2}$. We then consider the effect under the transformed Hamiltonian, where  $\hat H_{S_{r}}=e^{-\hat S_{r}}\hat He^{\hat S_{r}}=\omega\hat a^\dagger \hat a+\frac{\Omega}{2}\hat\sigma_z-G(\hat a^2+\hat a^{\dagger2})+\frac{\lambda^2}{\Omega}(1+\frac{\omega}{\Omega})\hat a^\dagger \hat a\hat\sigma_z-\frac{\lambda^2G}{\Omega^2}(\hat a^2+\hat a^{\dagger2})\hat\sigma_z-\frac{\lambda^4}{\Omega^3}\hat a^\dagger \hat a\hat a^\dagger \hat a \hat\sigma_z+\frac{\lambda^2}{\Omega}(1+\frac{\omega}{\Omega}-\frac{\lambda^2}{\Omega^2}-\frac{2\lambda^2}{\Omega^2}\hat a^\dagger \hat a)\vert\uparrow\rangle\langle\uparrow\vert+\mathcal{O}(\beta_i^{-3/2})$, for which the major contribution to the correction is about $\beta_i^{{-1}}\Delta^{-5/2}$.
We compare the above two influences and find that the major part is $\langle\hat{X}\rangle_{r}\simeq\beta_i^{{-1}}\Delta^{-5/2}$. Similarly, the correction of $(\Delta\hat{X})^2$ is on the order of $(\Delta \hat{X}_{r})^2\simeq\beta_i^{{-1}}\Delta^{-3}$. 
To ensure that our analysis above is valid, both $\langle\hat{X}\rangle_{r}$ and $(\Delta\hat{X}_{r})^2$ must be sufficiently small at the working point $t=\tau$. Such a requirement makes it necessary to restrict $\Delta\gg \beta_i^{{-1/3}}$ to make the higher-order corrections negligible, as verified by our numerical simulation of the inverted variance. As shown in Fig. \ref{ccc}(a), when the condition is satisfied, the performance of our solutions can be sustained. One key point that is of great importance is the precision of the sensing can be enhanced in virtue of the adjustable strength of the squeezing drive, which seems to essentially  relax the experimental requirement for the frequency ratio between the qubit and the field mode. This is strongly supported by the numerical outcomes illustrated in Fig. \ref{ccc}(b), showing the stringent requirement for the large $\beta_1$ to realize higher sensing precision can be  relaxed by appropriately reducing the  strength of the squeezing drive. Actually, there is a trade-off in between, as the improvement in precision by reducing the strength of the squeezing drive means requiring a greater coupling strength between the two-level system and the bosonic field.

\section{conclusion}
In summary, we have investigated quantum sensing based on critical phenomena of the SJCM. The two-photon drive enables the JCM to exhibit QRM-like dynamics and associated critical behaviors, without the requirement to reach the ultrastrong coupling regime. The model can be readily realized in different spin-boson systems. In an ion trap, the JC interaction between the internal and external degrees of freedom of a trapped ion can be mediated by a laser tuned to the first red sideband, while the squeezing driving can be realized with a Raman-type driving \cite{conclusion111}. For a circuit quantum electrodynamics architecture, the interaction between a superconducting qubit and a resonator is naturally described by the JCM, and the squeezing driving can be realized by a nonlinear process \cite{conclusion112,conclusion113}. These experimental advances make it possible to engineer the SJCM, and realize the critical dynamics for enhanced quantum sensing.
\section{acknowledge}
We thank Shaoliang Zhang and Huaizhi Wu for valuable discussions. This work was supported by the National Natural Science Foundation of China (Grants No. 12274080, No. 11874114, No. 11875108, No. 11705030), the National Youth Science Foundation of China (Grant No. 12204105), the Educational Research Project for Young and Middle-aged Teachers of Fujian Province (Grant No. JAT210041) and the Natural Science Foundation of Fujian Province (Grants No. 2021J01574 and No. 2022J05116).

\appendix\section{The detailed derivation of the quadrature dynamics of the SJCM} \label{A}
In the Heisenberg picture, the evolution of the quadrature operator can be written as 

\begin{eqnarray}
	\hat X(t)&=&\cos{(\sqrt{\Delta}t/2)}\hat X \nonumber\\
	&&+4(\alpha+2G)\Delta^{-\frac{1}{2}}\sin{(\sqrt\Delta t/2)}\hat P,
\end{eqnarray}%
where $\alpha=(\omega-2G)(1-g^2)/2$ and $g=\lambda/\sqrt{\Omega(\omega-2G)}$. For convenience, we choose the initial state of the bosonic field $\vert \varphi \rangle=(\vert 0\rangle+i\vert 1 \rangle)/\sqrt 2$ in the main text, and the mean value of operator $\langle \varphi\vert\hat X\vert\varphi\rangle=0$ and $\langle \varphi\vert\hat P\vert\varphi\rangle=1/\sqrt{2}$. The mean value of $\hat X(t)$ can be given by 

\begin{eqnarray}
	\langle\hat X\rangle_t=2\sqrt{2}(\alpha+2G)\Delta^{-\frac{1}{2}}\sin{(\sqrt{\Delta}t/2)},
\end{eqnarray}%
from which the susceptibility with the parameter $g$ can be obtained  as 
\begin{widetext}
	\begin{eqnarray}
		\chi_g(t)=\partial_g\langle\hat X\rangle_t &=&-2\sqrt{2}g(\omega-2G)\Delta^{-\frac{1}{2}}\sin{(\sqrt{\Delta}t/2)} 
		+32\sqrt{2}g(\omega-2G)(\alpha+G)(\alpha+2G)\Delta^{-\frac{3}{2}}\sin{(\sqrt{\Delta}t/2)} \nonumber\\
		&&-16\sqrt{2}g(\omega-2G)(\alpha+G)(\alpha+2G)\Delta^{-1}t\cos{(\sqrt{\Delta}t/2)}.
	\end{eqnarray}%
\end{widetext}
Similarly, the susceptibility of the bosonic field frequency $\omega$ can be expressed as 
\begin{figure}[!hbp]
		\includegraphics[width=0.5\textwidth]{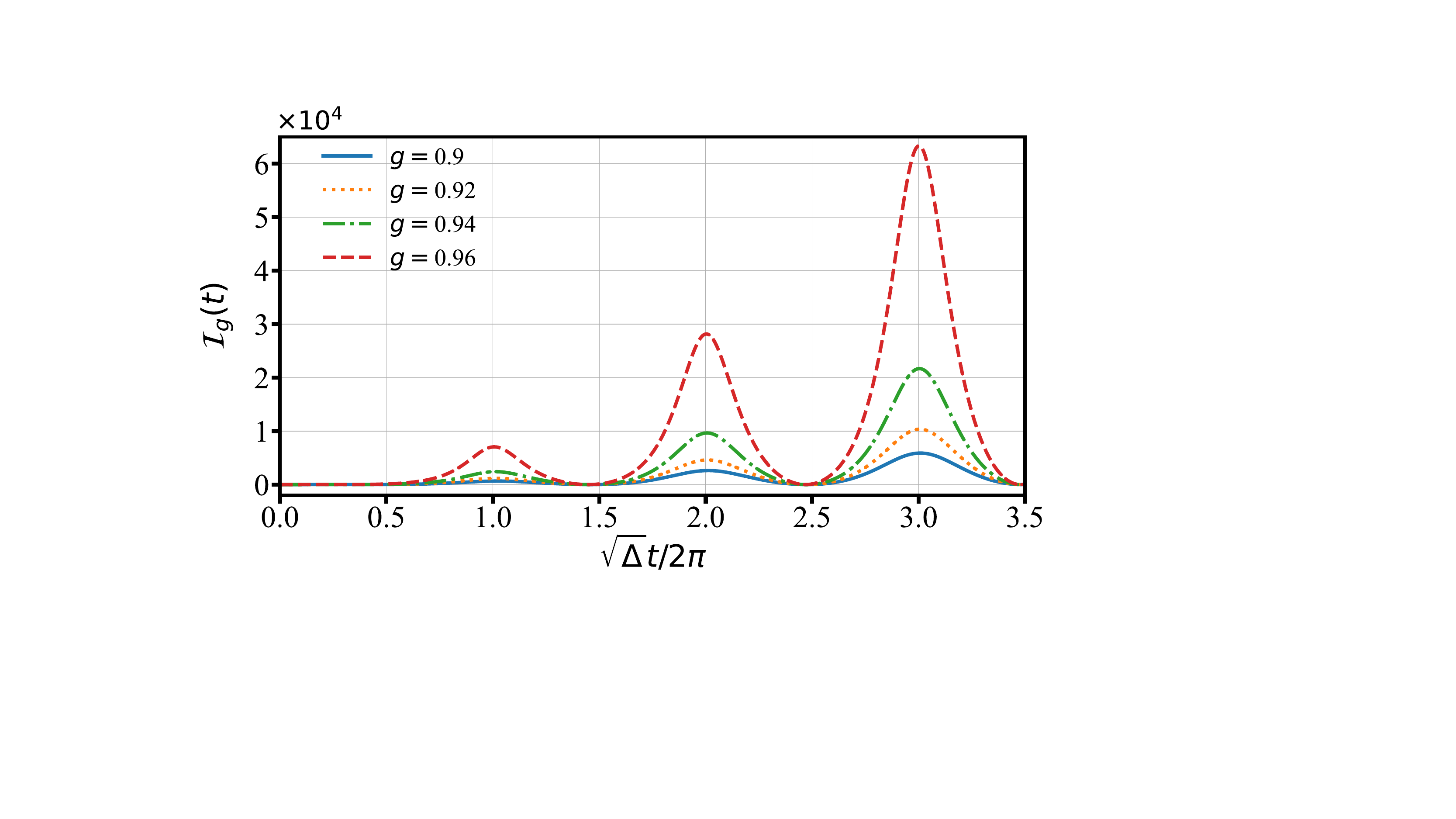}
	\caption{The inverted variance $\mathcal{I}_{g}(t)$ as a function of the evolution time $t$.}
	\label{ddd}
\end{figure}
\begin{eqnarray}
	\chi_{\omega}(t)=\partial_{\omega}\langle\hat X\rangle_t=-\frac{1}{2g(\omega-2G)}\chi_g(t).
\end{eqnarray}%
As shown above, both $\chi_g(t)$ and $\chi_{\omega}(t)$ exhibit the critical behaviors as $\Delta\to 0$. For $t=\tau_n=2n\pi/\sqrt\Delta(n\in{N}^+)$, we get
\begin{eqnarray}
	\chi_g(t)&=&(-1)^{n-1}16\sqrt{2}g(\omega-2G)\nonumber\\
	&&\times(\alpha+G)(\alpha+2G)\Delta^{-1}t,
\end{eqnarray}%
and 
\begin{eqnarray}
	\chi_{\omega}(t)=(-1)^{n}8\sqrt{2}(\alpha+G)(\alpha+2G)\Delta^{-1}t.
\end{eqnarray}%

We then calculate the variance of the quadrature operator $\hat X$ to determine the measurement precision of the 
correlated parameters. After a detailed calculation, we get $\langle \hat {X}^2\rangle_t=\cos^2(\sqrt{\Delta}t/2)+16(\alpha+2G)^2\Delta^{-1}\sin^2(\sqrt{\Delta}t/2)$, which leads to the variance
\begin{eqnarray}
	(\Delta\hat{X})^2&=&\cos^2(\sqrt{\Delta}t/2) \nonumber\\
	&&+8(\alpha+2G)^2\Delta^{-1}\sin^2(\sqrt{\Delta}t/2),
\end{eqnarray}%
reaching its minimums at $t=\tau_n=2n\pi/\sqrt\Delta$ $(n\in{N}^+)$. This result provides strong evidence
to the implementation of criticality-enhanced measurement precision of the correlated parameter $g$. As the corresponding inverted variance is defined as $\mathcal{I}_g(t)=\chi^2_g/(\Delta\hat{X})^2$, which achieves its local maximums $\mathcal{I}_g(\tau_n)= 2048n^2 \pi^2 (\omega-2G)^2g^{2}(\alpha+G)^2(\alpha+2G)^2\Delta^{-3}$ at $t=2n\pi/\sqrt\Delta$ $(n\in{N}^+)$ and  has a certain width at a large value over a period of the evolution time, as shown in Fig. \ref{ddd}. Similarly, we get $\mathcal{I}_{\omega}(\tau_n)=512n^2\pi^2(\alpha+G)^2(\alpha+2G)^2\Delta^{-3}$.

As compared to the Rabi model \cite{ref5}, it should be pointed out that the main advantage of the SJCM is reflected in the fact that it achieves the equivalent Rabi model through the addition of a two-photon-drive to the bosonic field of the JC interaction model, and effectively relaxes the dependence on the ultrastrong coupling by changing the boundary conditions of the phase transition. It was embodied by the ground state $\vert \phi_{np}^\downarrow\rangle=e^{r(a^2-a^{\dagger2})/2}\vert 0\rangle\vert g\rangle$ with a different compression parameter $r$. Based on this, the appendices are extended sections of Ref. \cite{ref21121} in terms of the encoding schemes, achieving the sensing with the same order of precision (see Fig. \ref{aaa}(b) in the main text), marking the feasibility of the SJCM in quantum sensing.
\section{Derivation of local observable of the two-level system of the SJCM}\label{B}

In the main text, we show that one can obtain the information of the parameter by measuring the two-level system directly \cite{ref21121}. It obtains the observable $\langle\hat\sigma_x\rangle=Re[\langle\varphi\vert u^{\dag}_{\uparrow}u_{\downarrow}\vert\varphi\rangle]$ if we set $2c^*_1c_2=1$ for the initial state of the two-level system. The inverted variance to estimate the precision of the parameter $g$ can be written as
\begin{eqnarray}
	\mathcal{I}_g=\frac{(\partial_g\langle\hat\sigma_x\rangle)^2}{1-\langle\hat\sigma_x\rangle^2}=\frac{Re[\partial_g \langle\varphi\vert u^{\dag}_{\uparrow}u_{\downarrow}\vert\varphi\rangle]^2}{1-Re[\langle\varphi\vert u^{\dag}_{\uparrow}u_{\downarrow}\vert\varphi\rangle]^2}.
\end{eqnarray}%
The initial state $\vert\varphi\rangle$ of the bosonic field in the eigenbasis of $\hat H^{\uparrow}_{np}=(\omega+\frac{\lambda^2}{\Omega})\hat a^{\dag}\hat a-G(\hat a^{\dag2}+\hat a^2)$ and $\hat H^{\downarrow}_{np}=(\omega-\frac{\lambda^2}{\Omega})\hat a^{\dag}\hat a-G(\hat a^{\dag2}+\hat a^2)$ can be expanded as

\begin{eqnarray}
	\vert\varphi\rangle=\sum_nc^\uparrow_n\vert\phi^\uparrow_n\rangle=\sum_nc^\downarrow_n\vert\phi^\downarrow_n\rangle,
\end{eqnarray}%
where $\vert\phi^\sigma_{n}\rangle=\hat S[r_\sigma]\vert n\rangle$, $\hat S[r_\sigma]=exp[-(r_{\sigma}/2)(\hat a^{\dag 2}-\hat a^2)]$ with $r_\downarrow=\frac{1}{4}\ln{\frac{\omega-\frac{\lambda^2}{\Omega}-2G}{\omega-\frac{\lambda^2}{\Omega}+2G}}$ and $r_\uparrow=\frac{1}{4}\ln{\frac{\omega+\frac{\lambda^2}{\Omega}-2G}{\omega+\frac{\lambda^2}{\Omega}+2G}}$. 
Then, we can get 
\begin{eqnarray}
	\langle\varphi\vert u^{\dag}_{\uparrow}u_{\downarrow}\vert\varphi\rangle&=&\langle\varphi\vert e^{i\hat H^\uparrow_{np}t}e^{-i\hat H^\downarrow_{np}t}\vert\varphi\rangle	\nonumber\\
	&=&\sum_{m,n}c^{\uparrow *}_mc^\downarrow_ne^{i(mE_\uparrow-nE_\downarrow)t}\langle\phi^\uparrow_m\vert\phi^\downarrow_n\rangle ,
\end{eqnarray}%
where $E_{\uparrow}=\sqrt{(\omega+\lambda^2/{\Omega})^2-4G^2}$ and $E_\downarrow=\sqrt{(\omega-{\lambda^2}/{\Omega})^2-4G^2}$.

Now if we choose the evolution time $\tau=4\pi/\sqrt{\Delta}$, we obtain

\begin{eqnarray}
	e^{-i\hat H^\downarrow_{np}\tau}\vert\varphi\rangle&=&\sum_{n}c^\downarrow_ne^{-in\sqrt{(\omega-{\lambda^2}/{\Omega})^2-4G^2})\tau}\vert\phi^\downarrow_n\rangle   \nonumber\\
	&=&\sum_{n}c^\downarrow_ne^{-i2\pi n}\vert\phi^\downarrow_n\rangle=\vert\varphi\rangle.
\end{eqnarray}%
At this time point, it has 
\begin{eqnarray}
	\langle\varphi\vert u^{\dag}_{\uparrow}u_{\downarrow}\vert\varphi\rangle&=&\langle\varphi\vert e^{i\hat H^\uparrow_{np}\tau}\vert\varphi\rangle  \nonumber\\
	&=&\sum_{m,n}c^{\uparrow *}_mc^\uparrow_ne^{im\tau\sqrt{(\omega+\lambda^2/{\Omega})^2-4G^2}}\langle\phi^\uparrow_m\vert\phi^\uparrow_n\rangle  \nonumber\\
	&=&\sum_{n}\vert{c^{\uparrow}_n}\vert^2e^{i2m\pi\mathcal{R}(g)},
\end{eqnarray}%
where $R(g)=\sqrt{\frac{(\omega+\lambda^2/{\Omega})^2-4G^2}{(\omega-\lambda^2/{\Omega})^2-4G^2}}$ and $\mathcal{R}(g)=R(g)-\lfloor R(g)\rfloor$. If $\mathcal{R}(g)=0.5$ is set, we get $\langle\varphi\vert u^{\dag}_{\uparrow}u_{\downarrow}\vert\varphi\rangle=\sum_n(-1)^n\vert c^\uparrow_n\vert^2$, which is approximately zero if the coefficient $c^\uparrow_n$ varies very slowly as $n$ increases. Therefore, we set the working point as $g=g_c$ that satisfies $\mathcal{R}(g_c)=0.5$ and the associated evolution time as $\tau=4\pi/\sqrt\Delta$. It can be seen that $\langle\hat\sigma_x\rangle\simeq0$ and $\partial_g \langle\varphi\vert u^{\dag}_{\uparrow}u_{\downarrow}\vert\varphi\rangle$
can be simplified as 
\begin{eqnarray}
	\partial_g \langle\varphi\vert u^{\dag}_{\uparrow}u_{\downarrow}\vert\varphi\rangle&=&[\langle\varphi\vert(\partial_\alpha u^\dag_\uparrow)u_\downarrow\vert\varphi\rangle+\langle\varphi\vert u^\dag_\uparrow(\partial_\alpha u_\downarrow)\vert\varphi\rangle]\frac{\partial\alpha}{\partial g} \nonumber\\
	&\simeq&-i\frac{\sin(\sqrt{\Delta}t)-\sqrt{\Delta}t}{\Delta^{\frac{3}{2}}}16Gg_c(\omega-2G) \nonumber\\ &&\times (\alpha+2G)\langle\varphi\vert u^{\dag}_{\uparrow}u_{\downarrow}\hat P^2\vert\varphi\rangle.
\end{eqnarray}%
The corresponding inverted variance is thus expressed as 

\begin{eqnarray}
	\mathcal{I}(g_c)&\simeq& Re[\partial_g \langle\varphi\vert u^{\dag}_{\uparrow}u_{\downarrow}\vert\varphi\rangle]^2 \nonumber\\
	&\simeq&4096\pi^2g_c^2G^2(\omega-2G)^2(\alpha+2G)^2\Delta^{-3}\nonumber\\
	&&\times (Im[\langle\varphi\vert u^{\dag}_{\uparrow}u_{\downarrow}\hat P^2\vert\varphi\rangle])^2,
\end{eqnarray}%
where $\langle\varphi\vert u^{\dag}_{\uparrow}u_{\downarrow}\hat P^2\vert\varphi\rangle$ is a constant independent of $g_c$. It can be seen that $\mathcal{I}(g_c)$ is comparable to the QFI we derived before (as shown in Fig. \ref{bbb} in the main text).

\bibliography{ref}

\end{document}